\documentstyle[aps,epsf,multicol]{revtex}

\def\avg#1{\left < {#1} \right >}
\def\etal{{\it et. al. }}
\def\prb{Phys. Rev. B }
\def\prl{Phys. Rev. Lett. }

\begin{document}

\draft
\title{ Quantum Stochastic Absorption }
\author{Sandeep K. Joshi\cite{jos}}
\address{Institute of Physics, Sachivalaya Marg, Bhubaneswar 751 005, India}

\author{Debendranath Sahoo\cite{dsahoo}\thanks{Present
address: Institute of Physics, Sachivalaya Marg, Bhubaneswar 751 005, India}}
\address{Materials Science Division, Indira Gandhi Centre for Atomic Research, 
Kalpakkam 603102, India} 

\author{A. M. Jayannavar\cite{amj} }
\address{Institute of Physics, Sachivalaya Marg, Bhubaneswar 751 005, India}

\maketitle

\begin{abstract} 

We report a detailed and systematic study of wave propagation through a
stochastic absorbing random medium. Stochastic absorption is modeled by
introducing an attenuation constant per unit length $\alpha$ in the free
propagation region of the one-dimensional disordered chain of delta
function scatterers. The average value of the logarithm of transmission
coefficient decreases linearly with the length of the sample. The
localization length is given by $\xi ~ = ~ \xi_w \xi_\alpha / ( \xi_w +
\xi_\alpha )$, where $\xi_w$ and $\xi_\alpha$ are the localization lengths
in the presence of only disorder and of only absorption respectively.
Absorption does not introduce any additional reflection in the limit of
large $\alpha$, i.e., reflection shows a monotonic decrease with $\alpha$
and tends to zero in the limit of $\alpha\rightarrow\infty$, in contrast
to the behavior observed in case of coherent absorption. The stationary
distribution of reflection coefficient agrees well with the analytical
results obtained within random phase approximation (RPA) in a larger
parameter space. We also emphasize the major differences between the
results of stochastic and coherent absorption.

\end{abstract}
\begin{multicols}{2}
\narrowtext

\section{Introduction}

Wave propagation in an active random medium has attracted much attention
during the past decade. Recently, many experiments have reported lasing
action of light in optically active strongly scattering media\cite{expt}.
These systems exhibit interesting physical properties due to the combined
effects of static disorder-induced multiple scattering and of coherent
amplification/absorption\cite{Nkupp,abhi,freil1,zhang,asen,pusti,%
misir,freil2,joshi1,joshi2,joshi3,jiang,beenak,pass}. In the extensively
studied case of an electron motion in a random medium it is well
established that quantum interference effects arising from a serial
disorder in one-dimensional systems lead to Anderson
localization\cite{tvr,nkuamj}. An essential difference between electron
and light propagation is the absence of conservation law for photons.
Light can be absorbed or amplified retaining the phase coherence. In most
of the theoretical studies amplification or absorption is modeled
phenomenologically by introducing an imaginary potential (optical
potential) in the Hamiltonian. In the case of light (electro-magnetic
waves) this corresponds to a medium with a complex dielectric constant.
Several interesting effects have been predicted which include statistics
of super-reflection and transmission\cite{Nkupp,abhi,freil1,zhang,asen,%
pusti,misir,freil2,joshi1,joshi2,joshi3,jiang} and the dual symmetry
between absorption and amplification\cite{beenak,pass}. Media thus modeled
are referred to as coherently absorbing or amplifying.

In the case of electron transport, inelastic scattering (due to phonons)
leads to loss of phase memory of the wave function. Thus the motion of
electrons becomes phase incoherent and sample to sample fluctuations
become self-averaging in the high temperature limit leading to a classical
behavior. There has been much interest in the effect of inelastic
scattering on the coherent tunneling through potential barriers.  To allow
for the possibility of inelastic decay on the otherwise coherent tunneling
through potential barriers, several studies invoke
absorption\cite{stone,zohta}. To study the above phenomenon, one resorts
to the optical potential models (coherent absorption models).

In the optical potential model the potential is made complex
$V(x)=V_r(x)-iV_i$. The Hamiltonian becomes non-Hermitian resulting in
absorption or amplification of probability current depending on the sign
of $V_i$. The presence of imaginary potential (absorption/amplification)
leads to many counter-intuitive features. In the scattering case, in the
vicinity of the absorber, the particle experiences a mismatch in the
potential (being complex) and therefore it tries to avoid this region by
enhanced back reflection. Imaginary potential plays a dual role of an
absorber and a reflector\cite{abhi,ruby}. In other words, in such models
absorption without reflection is not possible. Naively one expects the
absorption to increase monotonically as a function of $V_i$. However, the
observed behavior is non-monotonic\cite{abhi,ruby}. At first absorption
increases and after exhibiting a maximum decreases to zero as $V_i
\rightarrow \infty$. The absorber, in this limit acts as a perfect
reflector. During each scattering event an electron picks up an additional
scattering phase shift due to $V_i$ which along with multiple interference
leads to additional coherence or resonances in the system\cite{amj1}.
Thus, due to the presence of imaginary potentials, we have additional
reflection and resonances in the system. In the presence of coherent
absorption and quenched disorder, the stationary distribution for
reflection coefficient has been calculated\cite{Nkupp}. This has been done
within random phase approximation (RPA) using the invariant imbedding
method\cite{rrbd}. The stationary distribution is given by:

\begin{eqnarray}
\label{psr}
P_s(r) & = & \frac{|D|exp(|D|)exp(-\frac{|D|}{1-r})}{(1-r)^2} ~~\mbox{for}~~r \leq 1 \label{prad} \\
       & = & 0  ~~~~~~~~~~~~~~~~~~~~~~~~~~~~~~~\mbox{for}~~r > 1 .\nonumber
\end{eqnarray}

\noindent Here $D$ is proportional to $V_i/W$, $W$ being the strength of
disorder. Notice that the distribution has a single peak which shifts
towards $r=0$ with increasing absorption strength $V_i$. However, the
exact distribution obtained numerically for strong disorder and strong
absorption shows significant qualitative departure from this analytical
distribution\cite{abhi,jiang}. For sufficiently strong absorption, the
numerically obtained stationary distribution shows a double peak
structure. In the limit $V_i \rightarrow \infty$ the distribution becomes
a delta function at $r=1$. This corresponds to the limit where the
absorber acts as a perfect reflector.

To this end we would like to develop a model where absorption does not
lead to concomitant reflection and additional resonances as discussed
above. Recently, such a stochastic absorption model was developed by
Pradhan\cite{ppcm,pthesis} based on the work of B\"uttiker\cite{butt,hey}.
In his treatment several absorptive side-channels are added to the purely
elastic channels of interest. A particle that once enters the absorbing or
the side-channel never returns back and is physically lost. He has
obtained the Langevin equation for the reflection amplitude $R(L)$ for a
random medium of length $L$ by enlarging the $S$-matrix to include
side-channels. In continuum limit the equation for $R(L)$
is\cite{ppcm,pthesis}:

\begin{equation}
\label{langeq}
\frac{dR}{dL}=-\alpha R(L)+2ikR(L)+ikV(L)[1+R(L)]^2,
\end{equation}

\noindent where $\alpha$ is the absorption parameter and $V(L)$ is the
random potential representing the static disorder. Interestingly, within
the random phase approximation (RPA), the stationary probability
distribution for the reflection coefficient $P_s(r)$ ( for $L \rightarrow
\infty$ ) is again given\cite{ppcm,pthesis} by Eq.\ref{psr}. In our
present work we develop another simple model for absorption which can be
readily used to study the case of amplifying medium as well. The medium
comprises of random strength delta function scatterers at regular spatial
intervals $a$. To model absorption (leaking out) of electrons, an
attenuation constant per unit length $\alpha$ is introduced. Every time
the electron traverses the free region between the delta scatterers, we
insert a factor $exp(-\alpha a)$ in the free propagator following Ref.\
\onlinecite{datta}.  We find that this method of modeling absorption does
not lead to additional reflection and resonances as in the case of optical
potential models. We obtain the localization length and study the
statistics of reflection and transmission coefficients. The stationary
distribution of reflection coefficient agrees with Eq.\ref{psr} in a
larger parameter space. Following earlier method\cite{ppcm}, the continuum
limit of our model leads to the same Langevin equation (Eq. \ref{langeq})
for $R(L)$ where $\alpha$ is replaced by $2\alpha$. Naturally, agreement
of our result with Eq.\ref{psr} follows. In Sec.\ref{sec:model} we give
the details of our model and the numerical procedure. The section after
that is devoted to results and discussion.

\section{The Model}
\label{sec:model}

We carry out calculations on the wave propagation in an absorbing medium
characterized by an attenuation constant $\alpha$ and interspersed by a
chain of uniformly spaced independent delta-function scatterers of random
strengths. The $i^{th}$ delta-function scattering center is described by a
transfer matrix\cite{abrahams}

$$ {\sf M_i}~=~
\left ( \begin{array}{cc}
1-iq_i/2k & -iq_i/2k \\
iq_i/2k & 1+iq_i/2k \\
\end{array} \right ) $$

\noindent where $q_i$ is the strength of the $i^{th}$ delta-function. The
$q_i$'s are uniformly distributed over the range $-W/2 \leq q_i \leq W/2$,
i.e., $P(q_i)=1/W$. Here $W$ is the disorder strength. We set units of
$\hbar$ and $2m$ to be unity. The energy of the incident wave is $E=k^2$.
For further analysis, $W$ and $\alpha$ are scaled with respect to $a$ and
are made dimensionless. Propagation of the wave in-between two consecutive
delta-function scatterers separated by a unit spacing ($a=1$) can be
described by the matrix

$$ {\sf X}~=~
\left ( \begin{array}{cc}
e^{ik-\alpha} & 0 \\
0 & e^{-ik+\alpha} \\
\end{array} \right ). $$

The total transfer matrix for the $L$-site system is constructed by
repeated application of ${\sf M_i}$ and ${\sf X}$\cite{abrahams}: $$
M~=~{\sf M_LX....XM_2XM_1}. $$ From $M$ the reflection and transmission
amplitudes are calculated using $$ R~=~-\frac{M(2,1)}{M(2,2)} $$ and
$$T~=~-\frac{{\sf det}M}{M(2,2)}.$$ The reflection and transmission
coefficients are $r=|R|^2$ and $t=|T|^2$ respectively and the absorption
is given by $\sigma=1-r-t$. Thus, due to absorption the total flux is not
conserved and we have $r+t\neq1$.

\section{Results and Discussion}

In our studies we consider at least 10,000 realizations for calculating
various distributions and averages. In the case of stationary
distributions, the length of the samples considered were about 5 to 10
times the localization length. We also verified that the corresponding
distributions or averages do not evolve any further with increasing sample
length $L$. All results are shown for incident energy $E=k^2=1.0$ unless
specified otherwise.

\begin{figure}
\protect\centerline{\epsfxsize=2.5in \epsfbox{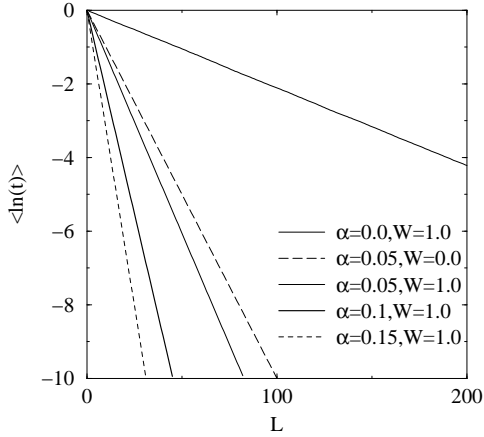}}
\caption{Average of logarithm of transmission coefficient $t$ versus
length $L$ for different values of absorption strength $\alpha$ and
disorder strength $W$ as indicated in the figure.}
\label{tvsl} 
\end{figure}

\begin{figure}
\protect\centerline{\epsfxsize=2.5in \epsfbox{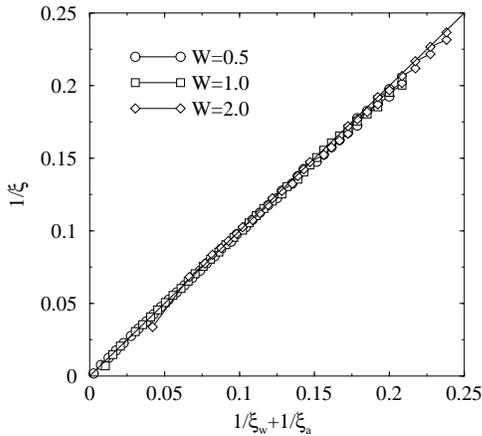}}
\caption{$1/\xi$ versus $1/\xi_w+1/\xi_a$, where $\xi_w$ is the
localization length for non-absorptive disordered system, $\xi_a$ is the
localization length for ordered absorptive system and $\xi$ is the
localization length for absorptive disordered system. }
\label{xifig}
\end{figure}

We first consider the behavior of $\avg{lnt}$. The angular bracket denotes
the ensemble average. In Fig.\ref{tvsl} we plot $\avg{lnt}$ as a function
of length $L$ for ordered absorptive medium ( $W=0.0$, $\alpha=0.05$ ),
ensemble averaged disordered non-absorptive medium ( $W=1.0$, $\alpha=0.0$
) and disordered absorptive medium ( $W=1.0$, $\alpha=0.05, 0.1, 0.15$ ).  
In all the cases transmission decays exponentially with the length. The
absorption-induced length scale $\xi$ in random medium associated with the
decay of transmission coefficient is always less than both $\xi_a$ and
$\xi_w$. The localization length for disordered nonabsorptive medium
scales as\cite{souk} $\xi_w=96k^2/W^2$ and for ordered absorptive medium,
as $\xi_a=1/\alpha$. In Fig.\ref{xifig} we show the plot of $1/\xi$ versus
$1/\xi_w+1/\xi_a$ obtained by changing $\alpha$ for various values of
disorder strength $W$. We have numerically calculated $1/\xi$ for the
different cases. All the points fall on a straight line with unit slope
indicating the relation $1/\xi~=~1/\xi_w+1/\xi_a$. Such a relation exists
for the case of coherently absorbing and amplifying
media\cite{zhang,jiang}.

\begin{figure}
\protect\centerline{\epsfxsize=2.5in \epsfbox{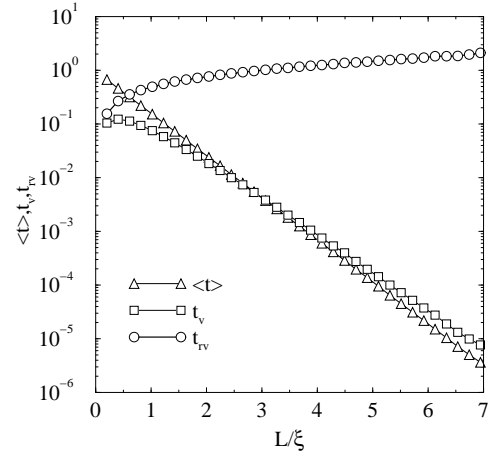}}
\caption{Average transmission coefficient $\avg{t}$, variance of $t$
($t_v$) and relative variance of $t$ ($t_{rv}$) as a function of $L/\xi$
for fixed disorder $W=1.0$ and absorption $\alpha=0.01$.}
\label{variance}
\end{figure}

To study the nature of fluctuations in the transmission coefficient, in
Fig.\ref{variance} we plot, on log-scale, average $t$ ($\avg{t}$),
root-mean-squared variance $t_v=\sqrt{\avg{t^2}-\avg{t}^2}$ and
root-mean-squared relative variance $t_{rv}=t_v/\avg{t}$ as a function of
length $L$ for $W=1.0$ and $\alpha=0.01$. We see from the figure that
$t_v$ is less than $\avg{t}$ and $t_{rv}$ is less than unity upto $L/\xi
\approx 3$. But, beyond that $t_v$ becomes greater than $\avg{t}$ and
$t_{rv}$ crosses unity making the transmission coefficient a
non-self-averaging quantity. This implies that the fluctuations in $t$
over the ensemble of macroscopically identical samples dominates the
ensemble average. The transmission coefficient becomes very sensitive to
spatial realizations of the impurity configurations.

\begin{figure}
\protect\centerline{\epsfxsize=2.5in \epsfbox{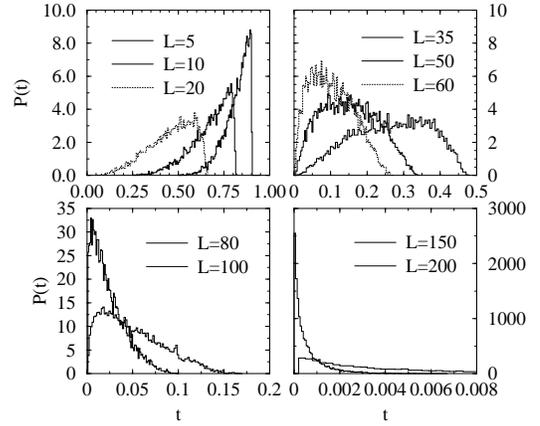}}
\caption{Distribution of transmission coefficient $t$ from a disordered
absorptive system with $W=1.0$ and $\alpha=0.01$ at different lengths
$L$.}
\label{tdist}
\end{figure}

In Fig.\ref{tdist} we show the distribution $P(t)$ at different sample
lengths for $\alpha=0.01$ and $W=1.0$. For small lengths $L$, resonant
transmission dominates and $P(t)$ peaks at a large value of $t$. In fact
for $L\rightarrow 0$, $P(t)\rightarrow\delta (t-1)$. As the length becomes
comparable to the localization length $L\sim\xi$, multiple reflections
start dominating. Consequently, the time spent inside the medium increases
leading to more absorption. Thus, the peak of the distribution shifts to
smaller values of $t$ and the distribution broadens due to randomization
by disorder. In the long length limit $L>>\xi$, the distribution develops
long tails and its peak shifts towards $t=0$. The transmittance shows
large sample-to-sample fluctuations and becomes a non-self-averaging
quantity. It is the tail of the distribution that determines the behavior
of fluctuations.  Finally, as expected, for $L\rightarrow \infty$ , $P(t)
\rightarrow \delta(t)$.

\begin{figure}
\protect\centerline{\epsfxsize=2.5in \epsfbox{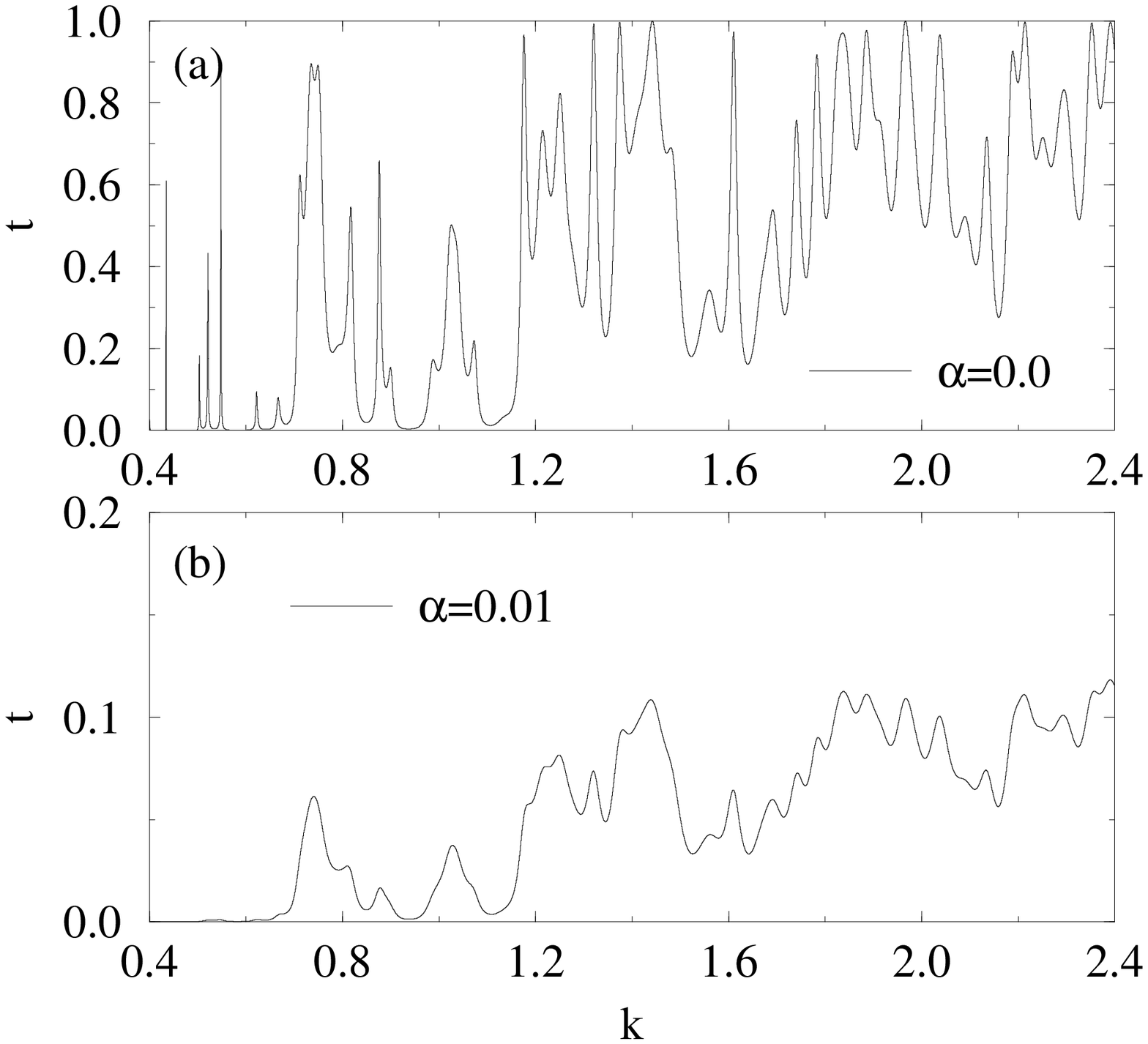}}
\protect\centerline{\epsfxsize=2.5in \epsfbox{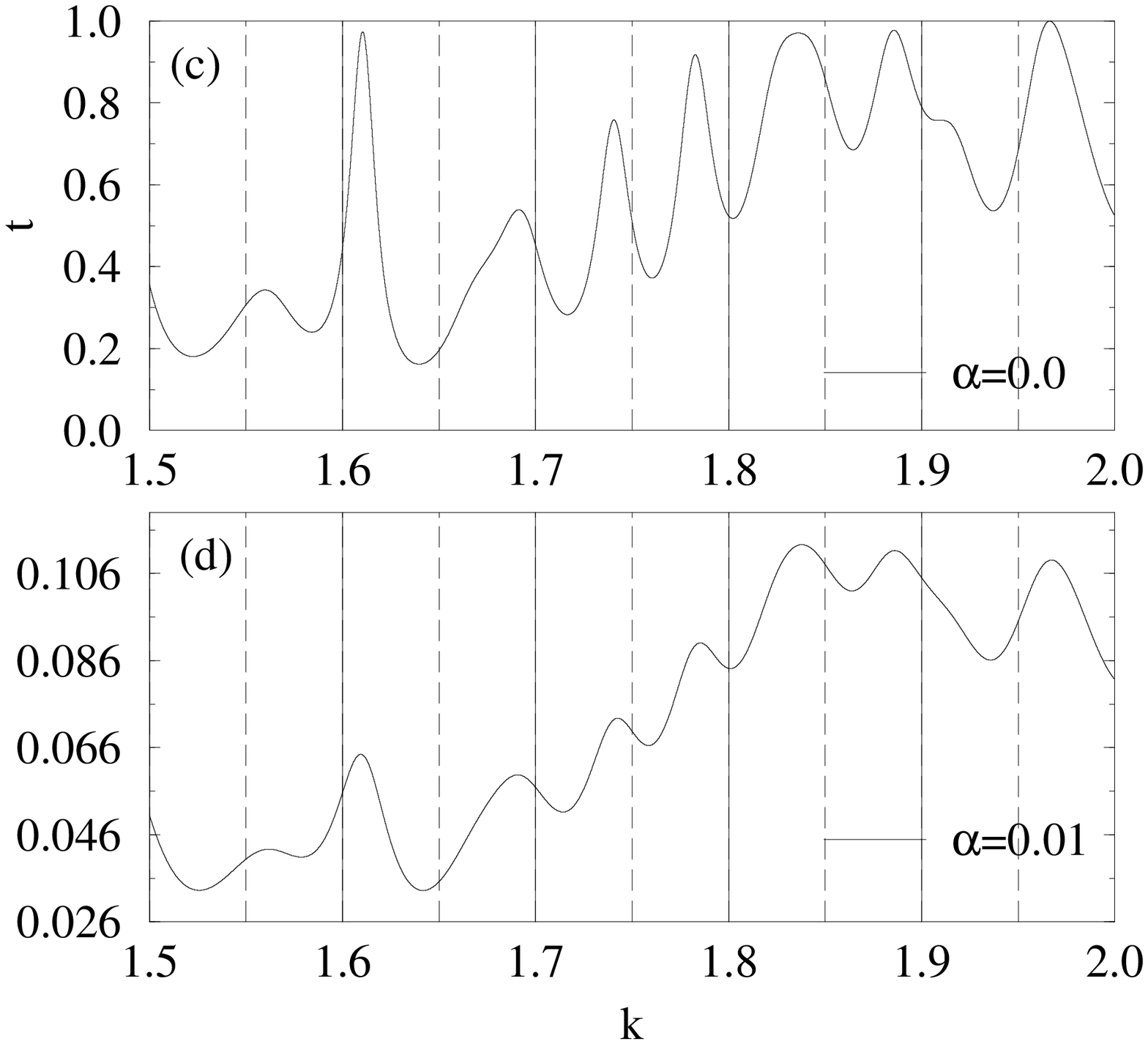}}
\caption{Transmission coefficient as a function of incident wavenumber $k$
for (a),(c) disordered non-absorptive sample $W=1.0$ of length $L=100$ and
(b),(d)  disordered absorptive sample $W=1.0,\alpha=0.01$ of length
$L=100$. }
\label{tvse}
\end{figure}

From the previous discussion it is clear that the transmittance becomes
non-self-averaging due to the appearance of tails in the $t$-distribution.  
These tails owe their existence to the presence of resonant realizations
in the ensemble. Hence, it is worthwhile to investigate the nature of
resonances and the effect of absorption on them. Specifically, we would
like to understand if the presence of absorption would give rise to any
new resonances. It is well known from the studies in passive disordered
media that the ensemble fluctuation and the fluctuations for a given
sample as a function of chemical potential or energy are expected to be
related by some sort of ergodicity\cite{ergod}, i.e., the measured
fluctuations as a function of the control parameter are identical to the
fluctuations observable by changing the impurity configurations. In
Fig.\ref{tvse}(a) we show the plot of $t$ versus $k$ for $W=1.0$ and
$\alpha=0$ at $L=100$ for a given realization of the random potential.
Figure \ref{tvse}(b) shows a plot of $t$ versus $k$ for the same
realization but with $\alpha=0.01$. By mere visual inspection one can see
that the only effect of absorption, apart from reducing the value of
transmission, is to increase the width of resonance peaks for the passive
case. Thus the presence of absorption does not introduce any new
resonances. This can be seen from Fig.\ref{tvse}(c) and (d) which
emphasize Fig.\ref{tvse}(a) and (b) respectively by enlarging a narrow
region between $k=1.5$ to $k=2.0$. We do not see any new peaks in the
transmission spectrum for absorptive case. Similar effect is observed in
case of reflection also.

\begin{figure}
\protect\centerline{\epsfxsize=2.5in \epsfbox{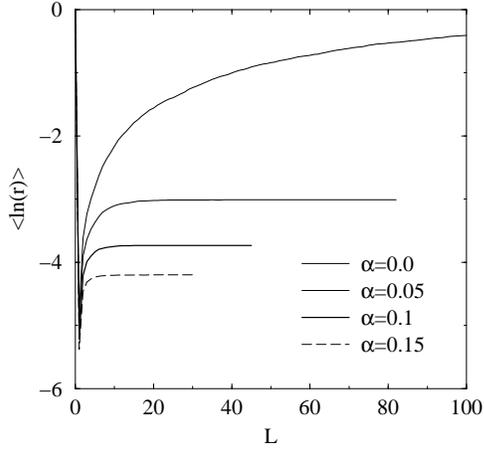}}
\caption{Average of logarithm of reflection coefficient versus sample
length for a fixed value of disorder $W=1.0$ and different values of
absorption $\alpha$ indicated in the figure. }
\label{rvsl}
\end{figure}

We now turn our attention to the statistics of reflection coefficient. In
Fig.\ref{rvsl} we plot $\avg{lnr}$ as a function of length $L$ for a fixed
value of disorder strength $W=1.0$ and different values of absorption
strength $\alpha$ as indicated in the figure. It increases with $L$
initially and for $L \gg \xi$, it saturates. At any $L$,
$\avg{lnr}|_{W,\alpha} < \avg{lnr}|_{W,0}$. This is in contrast to the
behavior observed for the case of coherent absorption. As we know, in the
case of coherent absorption, the reflection coefficient tends to unity for
absorption strength becoming very large. In this regime, the predominantly
reflecting nature of optical potential makes reflection larger than that
in the corresponding passive case.

\begin{figure}
\protect\centerline{\epsfxsize=2.5in \epsfbox{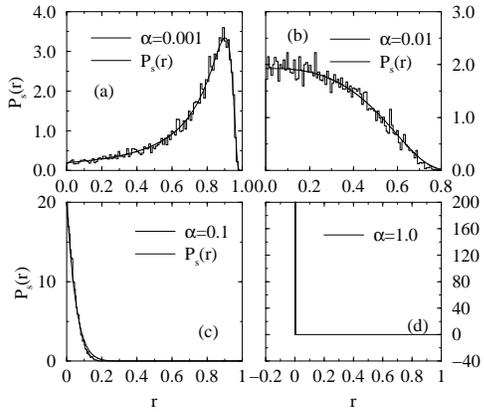}}
\caption{Stationary distribution of reflection coefficient for a fixed
value of disorder strength $W=1.0$ and different values of absorption
$\alpha$. The thick line shows the single parameter fit of analytical
expression Eqn.\ref{psr} with (a) $D=0.197$ for $\alpha=0.001$,(b)  
$D=1.92$ for $\alpha=0.01$ and (c) $D=20.53$ for $\alpha=0.1$.}
\label{rdist}
\end{figure}

\begin{figure}
\protect\centerline{\epsfxsize=2.5in \epsfbox{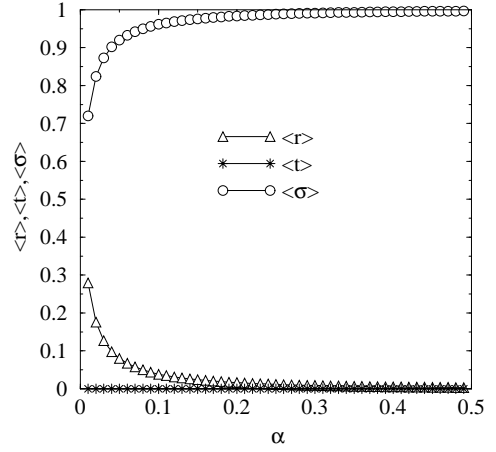}}
\caption{Average value of reflection coefficient ($\avg{r}$), transmission
coefficient ($\avg{t}$) and absorption ($\avg{\sigma}$) versus absorption
strength $\alpha$ for $W=1.0$ and $L/\xi=10$.}
\label{avgrvsalpha}
\end{figure}

In Fig.\ref{rdist} we have shown $P_s(r)$ for various values of $\alpha$.
In the small $\alpha$ range, i.e., for $\alpha=0.001$, the distribution
has a peak at large $r$. As we increase $\alpha$ the peak shifts to
smaller values of $r$. The thick line shows the fit obtained using the
analytical expression given in Eq.\ref{psr}.  In the limit of large
$\alpha$, the distribution tends to become a delta function at $r=0$. This
is in sharp contrast to the behavior observed for coherent
absorption\cite{abhi}. The distribution is always single peaked. For all
non-zero values of $\alpha$ the medium acts as an absorber only and there
is no additional reflection due to absorption. Figure \ref{avgrvsalpha}
shows a monotonic decrease of saturated value of average reflection
coefficient $\avg{r}_s$ as a function of $\alpha$. The average absorption,
defined as $\avg{\sigma}~=~1-\avg{r}-\avg{t}$, increases monotonically
with increasing $\alpha$ and in the limit of $\alpha \rightarrow \infty$
saturates to unity in contrast to the optical model wherein it tends to
zero. In the case of the optical model, the absorption coefficient is a
non-monotonic function of absorption strength $V_i$ and for values of
$V_i$ near the peak the stationary distribution of reflection coefficient
displays a double peak\cite{abhi,jiang}.  In fact our model exhibits the
properties in agreement with physical expectations of an absorbing medium,
i.e., stronger the absorption lesser are the reflection and transmission
across the medium.

\begin{figure}
\protect\centerline{\epsfxsize=2.5in \epsfbox{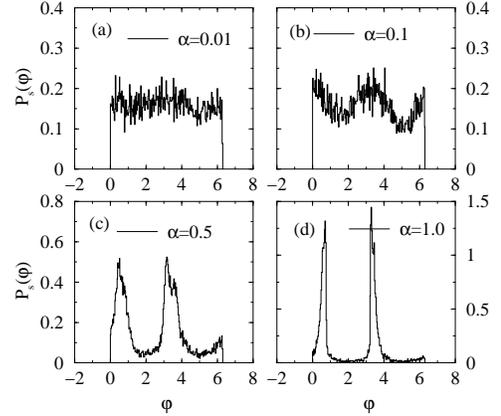}}
\caption{Stationary distribution of phase of reflected wave for fixed
disorder strength $W=1.0$ and different values of absorption $\alpha$.}
\label{pdist}
\end{figure}

Finally, we discuss the phase distribution. Figure \ref{pdist} shows the
stationary distribution of phase of the reflected wave for a fixed
disorder strength $W=1.0$ and various for values of $\alpha$. For small
values of disorder one generally expects the phase distribution to be
uniform if the system size is around the localization length. This is seen
in Fig.\ref{pdist}(a) for the case of weak absorption. As we increase
$\alpha$ the phase distribution develops two distinct peaks -- a feature
observed for coherent absorption also. This is related to the fact that
the localization length decreases with $\alpha$. We would like to point
out that the stationary distribution $P_s(r)$ is same within the RPA for
the case of coherently as well as stochastically absorbing media. Unlike
in the case of coherently absorbing medium, Eq.\ref{psr} seems to be valid
in a larger parameter space for the stochastic absorbing medium where the
RPA may not be valid. The parameters for validity of the RPA are
determined by the observation of uniform phase distribution. However,
beyond the RPA, $P_s(r)$ for the case of stochastic and coherent absorbing
media are qualitatively distinct from each other.

In conclusion, we have studied a new model of quantum stochastic
absorption. The behavior observed for transmission and reflection
coefficients is in accordance with physical expectations of an absorbing
medium. This model can be extended to the case of stochastically
amplifying medium. It exhibits duality between absorption and
amplification which has received much attention recently. Results for this
will be reported elsewhere\cite{xxx}.

\acknowledgments

One of us (DS) would like to thank Prof. S. N. Behera for extending
hospitality at the Institute of Physics, Bhubaneswar.

\end{multicols}
\end{document}